\documentclass[10pt,preprint]{aastex}

\newcommand {\ha} {H$\alpha$\,\,}
\newcommand {\kms} {\,km\,s$^{-1}$\,}

\newcommand {\has} {H$\alpha$}
\newcommand{\BB}{\textit{\texttt{B}\texttt{\has}\texttt{BAR}\,\,}}
\newcommand{\FM}{\texttt{\textsc{FaNTOmM}}}

\newcommand{\myemail}{ \it olivier@astro.umontreal.ca}
\newcommand{\GF}{{\sc GH$\alpha$FaS}}
\newcommand {\Msol} {M$_\odot$}

\slugcomment{To appear in PASP}

\shorttitle{\GF}
\shortauthors{Hernandez et al.}

\begin{document}


\title{\GF: Galaxy H-alpha Fabry-Perot System for the WHT}



\author{O. Hernandez\altaffilmark{1},
K. Fathi\altaffilmark{2,6},C. Carignan\altaffilmark{1}, J.
Beckman\altaffilmark{2,7}, J.-L. Gach\altaffilmark{3}, P.
Balard\altaffilmark{3} \\ and \\ P. Amram\altaffilmark{3}, J.
Boulesteix\altaffilmark{3}, R.L.M. Corradi\altaffilmark{8,2}, M-M. de
Denus-Baillargeon\altaffilmark{1,9}, B. Epinat\altaffilmark{3},M.
Rela\~no\altaffilmark{5}, S. Thibault\altaffilmark{4}, P.
Vall\'ee\altaffilmark{1} }

\altaffiltext{1}{Laboratoire d'Astrophysique Exp\'erimentale,
Observatoire du mont M\'egantic \& D\'epartement de physique,
Universit\'e de Montr\'eal, C.P. 6128 succ. centre ville,
Montr\'eal, Qu\'ebec, Canada H3C 3J7: \myemail}
\altaffiltext{2}{Instituto de Astrof\'\i sica de Canarias, C/ V\'\i
a L\'actea s/n, 38200 La Laguna, Tenerife, Spain}
\altaffiltext{3}{Laboratoire d'Astrophysique de Marseille,
Observatoire Astronomique Marseille-Provence, Universit\'e de
Provence \& CNRS, 2 place Le Verrier, 13248 Marseille Cedex 4,
France.} \altaffiltext{4}{IMMERVISION, 2020 University, suite 2420,
Montr\'eal, Qu\'e., Canada H3A 2A5} \altaffiltext{5}{Dpto. F\'isica
Te\'orica y del Cosmos, Universidad de Granada, Avda. Fuentenueva s/n,
18071 Granada, Spain} \altaffiltext{6}{Stockholm Observatory,
AlbaNova University Center, 106 91 Stockholm, Sweden}
\altaffiltext{7}{Consejo Superior de Investigaciones Cient\'\i
ficas, Spain} \altaffiltext{8}{Isaac Newton Group, Apartado de
Correos 321,   38700 Santa Cruz de la Palma, Spain} \altaffiltext{9}{
Institut Fresnel, CNRS \& Universit\'es Aix Marseille, 13397
Marseille Cedex 20, France}




\begin{abstract}

\GF\, a new Fabry-Perot system, is now available at the William
Herschel Telescope. It was mounted, for the first time, at the
Nasmyth focus of the 4.2 meter WHT on La Palma in July 2007. Using
modern technology, with a spectral resolution of the order
R$\sim$15000 , and with a seeing limited spatial resolution, \GF\
will provide a new look at the \has -emitting gas over a 4.8
arcminutes circular field in the nearby universe. Many science
programs can be done on a 4.2 metre class telescope in world class
seeing conditions with a scanning Fabry-Perot. Not only galaxies but
HII regions, planetary nebulae, supernova remnants and the diffuse
interstellar medium are subjects for which unique data can be
aquired rapidly. Astronomers from the Laboratoire d'Astrophysique
Exp\'erimentale (LAE) in Montr\'eal, the Laboratoire
d'Astrophysique de Marseille (LAM-OAMP), and the Instituto de
Astrof\'isica de Canarias (IAC),  have inaugurated \GF\ by studying
in detail the dynamics of some nearby spiral galaxies. A robust set of
state-of-the-arts tools for reducing and analyzing the data cubes
obtained with \GF\ has also been developed.
\end{abstract}



\section{Introduction}

From 2007 July 2nd to July 8th, the new private instrument \GF\ \footnote{www.astro.umontreal.ca/ghafas or www.iac.es/project/ghafas} was
commissioned on the Whilliam Herschel Teslecope (WHT). The acronym \GF\, stands for Galaxy Halpha
Fabry-Perot (FP) System, and gives an idea of the nature and the
prime use of the instrument. It is a new generation FP
interferometer, whose chief and powerful advantage is its high
sensitivity photon counting detector coupled to a large field of
view. \GF\, (which sounds like the Spanish word, "gafas", for a pair
of glasses) is used on the GHRIL platform of the Nasmyth focus of
the WHT.

\GF\, is an improved version of the scanning Fabry-Perot system
FaNTOmM \citep{he2003}, which is a resident instrument at the
Observatoire du mont M\'egantic (OmM) 1.6 m telescope and which has
also been used as a visitor instrument on the Canada-France-Hawaii
(CFH) and ESO La Silla 3.6 m telescopes. The complete system is
composed of a focal reducer, a calibration unit, a filter wheel for
the order sorter filters, an FP etalon and a photon-counting (IPCS)
camera. The IPCS is composed of an Hamamatsu intensifier
multi-channel plates (MCP) tube which intensifies every generated
electron coming from the photocatode by a factor 10$^7$. Each photon
event, recorded on a DALSA CCD, is then analyzed by a centering
algorithm. With this amplification, the camera has essentially no
readout noise. Because of this, a zero noise IPCS is to be preferred
to CCDs at very low flux level \citep{ga2002}, even if the GaAs IPCS
has only a DQE of $\sim$26\%. Moreover, because of the fast scanning
capability, it can average out the variation of atmospheric
transmission which is not possible with the long integration times
needed per channel for the CCDs in order to beat the read-out noise.

In the last 3 years, around 150 galaxies were observed with the
FaNTOmM system on the OMM, CFH and ESO La Silla telescopes in the
context of 3 large surveys: the SINGS samples (Daigle et al. 2006
\& Dicaire et al. 2007), a survey of barred galaxies, the \BB
sample (Hernandez et al. 2005) and a sample of Virgo spirals
(Chemin et al. 2006). While the first scientific justification was to derive high spatial resolution optical
rotation curves for mass modeling purposes, the data was also used
by IAC astronomers to constrain the role of gravitational
perturbations as well as feedback from individual HII regions on the
evolution of structures in galaxies \citep{2007ApJ...667L.137F,
2007A&A...467.1117R} and by a Berkeley-Munich group, the G\'EPI
group at Observatoire de Paris and Laboratoire d'Astrophysique de
Marseille (LAM) to compare those local samples to high z galaxies.

\GF\ comes with its own custom designed focal reducer developed to
be optically and mechanically compatible with the Nasmyth focus of
the WHT. The system has its own control and data acquisition system.
It has a 4 arcmin circular field with a 0.4 arcsec pixel and a
minimum of $\sim$5 km/s velocity resolution. Full acquisition and
reduction software (mainly based on IDL routines) is provided by the
Montr\'eal and IAC groups. The project will be done in 3 phases. For
Phase I (July 2007), the optical system (focal reducer, filter wheel
\& calibration unit) has been delivered to the WHT and used with the
Marseilles IPCS camera for this first run. For Phase II (beginning
of 2008), an improved GaAs IPCS will be added to the system. Phase
III (mid 2008) will provide an FP controller to replace the old
CS100 technology (provided by IC Optical formerly Queen's Gate) and
possibly a Tunable Source to calibrate the data at the observing
wavelength.

\GF\ is presented in this paper. Section 2 describes the science
cases for the \GF\ instrument, while Section 3 discusses the
technical details of the instrument.  In Section 4, the data
preparation and reduction procedures are presented while a sample of
the first results is shown. The conclusions are given in Section 5.

\section{\GF\ Science Drivers}
Several instruments exist that provide two dimensional maps for
kinematical and population analysis of galaxies or other
astronomical objects. Those observations are mainly based on two
different techniques of wavelength separation: dispersive or
scanning. The former uses fibers (\textsc{Virus}: Hill et al. 2006, \textsc{Argus}: Kaufer et al.2003, \textsc{Integral}: Arribas et al. 1998), micro-lenses (first
concept done by Courtes (1982), \textsc{Tiger}: Bacon et al. 1995,
\textsc{Sauron} Bacon et al. 2001), image slicers (\textsc{SINFONI}: Eisenhauer et al. 2003, \textsc{MUSE}:
Henault et al. 2004, \textsc{WiFes}: Dopita et al. 2007,
\textsc{SWIFT}: Thatte et al. 2006, \textsc{FRIDA}: L{\'o}pez et al.
2006, and references therein)  and the later Fabry-Perot
etalons in a pupil (\FM\ in Hernandez et al. 2003, \textsc{Cigale}:
Boulesteix et al. 1984, \textsc(HIFI): Bland et al. 1990,
\textsc{Taurus}: Atherton et al. 1982. \GF\, is a scanning
instrument with the largest field-of-view (FOV) with such a spatial
resolution (0.4"/px in low mode and 0.2"/px in high spatial 
resolution mode) on a 4m class telescope.

Two dimensional kinematics is a very powerful technique for studying
the structure and evolution of extended sources. Galaxy dynamics,
the distribution of dark matter, the dynamical and physical state of
star forming regions, circum nuclear starburst and fueling of active
galactic nuclei, detection of kinematically decoupled components,
dynamics and physical state of planetary nebulae, and
supernova-driven winds and feed-back mechanisms are just a few
phenomena which can be studied with this technique.

Large spectral-range spectrographs are commonly used to analyze the
stellar population and stellar and gaseous kinematics of relatively small
(1 kpc) regions of nearby galaxies or entire galaxies at high redshifts
(e.g., de Zeeuw et al. 2002, Genzel et al. 2006, Peletier et al. 2007, 
Emsellem et al. 2007, Marquart et al. 2007).
The low-redshift studies have shown that morphological methods are
not a sufficient tool for galaxy classification and supplementary
kinematical and population data are necessary for understanding
galaxy evolution. High-redshift, two-dimensional kinematical studies
have shown that galaxy disks are more disturbed at z $\sim$ 1 than
locally.

Large spatial-coverage instruments have been successful at studying
the state and dynamics of galaxies and of the interstellar medium
(e.g., Zurita et al. 2004, Hernandez et al. 2005a, Chemin et al.
2006, Sluis \& Williams 2006, Daigle et al. 2006a, Fathi et al. 2007a,
Rela\~no et al. 2005;2007, Rosado et al. 2007, Amram et al. 2007). These
studies have observationally confirmed the important role of
gravitational perturbations on evolution of galaxies, which are now
accepted to be the main drivers for morphological transformations.

\subsection{HII Region Studies and Interaction of Stellar Winds with the ISM}
The internal kinematics of HII regions are of interest because of
the light shed on the evolutionary properties of the massive stars
which ionize them and on their interaction with the surrounding ISM.
A key issue here is that the velocity dispersion measured in \ha of
the majority of HII regions observed in external galaxies, which are
necessarily selected as the most luminous regions, is found to be
supersonic. A basic question is what is the source of input energy
to maintain these line widths, and the linked possibility that the
line widths could be used as a luminosity index and hence as a
distance measuring standard candle. 
A long term debate was opened by Terlevich \& Melnick (1981) who
suggested that the line widths vary as the fourth power of the
luminosity and that this reflected virial equilibrium within the
regions. This was addressed in some detail by Chu \& Kennicutt
(1994a) in a study of 30 Doradus, and by Chu \& Kennicut (1994b), 
who found for a large sample of bright HII regions that their line 
profiles are broadened by turbulence as well as wind-induced line 
splittings. i.e. that in general virialization cannot be assumed. 
These studies are merely pointers to the wealth of physical
inference which will be available using \GF\, to investigate 
how metals are distributed around galaxies by the
outflows from HII regions and by similar flows around galactic
centres, and how the ionization structure of HII regions relates 
to their kinematic structure.

\subsection{Planetary Nebulae and Ionized Stellar Outflows}
In the last two decades, the extraordinary variety of morphologies
of planetary nebulae (PNe) has been widely recognized (e.g.: Balick
\& Frank 2002). Their study is crucial in order to fully understand
mass loss, the physical process that governs the latest evolution of
low mass stars and plays a basic role in the chemical enrichment of
galaxies. A precise determination of the 6-D geometry and kinematics of PNe
requires 2-D spectroscopy, at a resolution R$\ge$30000 for
``normal'' PNe (and lower for the highly-collimated, high-velocity
components) and in the brightest nebular emission lines (
[OIII], [H$\alpha$], and [NII]). \GF\, ca provide an efficient and accurate way to observe all the
different macro and microstructures present in the nebulae, which
include - often in the same object - multiple spherical shells,
bipolar or multipolar lobes, point-symmetrical features, symmetric
knots, filaments and jets.  Comparison of the images and radial
velocity fields with sophisticated radiative-hydrodynamical
simulations (e.g. Garc{\'{\i}}a-Segura et al. 1999) can then
reveal the physical processes at work in shaping PNe. 
By obtaining fully monochromatic images of the nebulae in different
emission lines, we can determine the density, temperature, and ionization degree,
throughout the objects. This is also essential for a proper 3-D
modelling of the nebulae and a correct determination of their
physico-chemical properties (cf. Gon{\c c}alves et al. 2006). Ultimately, the
most important questions to be answered are the following ones: why do
stars loose their spherical symmetry at the tip of the asymptotic
giant branch or immediately after?  Can single stars do it, or is the
majority of the observed PNe the result of binary evolution (e.g., Moe
\& de Marco 2007)?

\subsection{Structure and Evolution of Nearby Spiral Galaxies}
The evolution of structures in galaxies is given by the interactions
with neighboring galaxies, gravitational perturbations, and/or winds
from star forming regions. The IAC, LAE and LAM groups have developed a set of
robust data analysis tools which were used successfully on previous
Fabry-Perot observations covering out to 12 kpc in the nearby
grand-design Sc spiral galaxy, M~74. These tools are optimized to
investigate the large-scale dynamics as well as feedback from
individual HII regions into their surrounding interstellar medium in
order to study the evolution of structures in galaxies using
two-dimensional kinematical observations (Zurita et al. 2001, 
Hernandez et al. 2005a,b, Fathi et al. 2005,2007a,b).
After the careful examination of the resonance structures by
measuring the pattern speeds of the bars in M~100 (Hernandez et al.
2005b), NGC~1068 (Emsellem et al. 2006), and NGC~6946 (Fathi et al.
2007b), there is now a solid framework for kinematically studying the
resonant interaction in disks of galaxies. 
The combination of these studies with the emerging theoretical foundation 
(Maciejewski 2006, Maciejewski \& Athanassoula 2007, Athanassoula 2005) 
has opened the possibility to add the global kinematical information to
the local effects at or near the Lindblad and ultraharmonic
resonance radii, and compare the strength and degree of symmetry of
the density waves. Such quantification is important to study to what
degree bars infer dynamical mixing of the gas and stars in their parent disks.

\subsection{Fueling of Active Galactic Nuclei}
Due to the mass concentration towards the center of the
gravitational potential in galaxies, their nuclear region are
suitable environments for the formation of Super Massive Black Holes
(SMBH) which can become active by infall of interstellar gas. 
Recent works (Tremaine et al. 2002 and references therein) show 
in addition that the mass of the SMBH is proportional
to that of the bulge of its host galaxy. This relation suggests a
connection between the large-scale properties of galaxies and the
nuclear properties (Haehnelt \& Kauffmann 2000). The process
responsible for such a relation has to transfer matter and angular
momentum within galaxies. Pioneer works have proposed that bars can
trigger the mass flows to help grow and feed SMBHs in galaxies
(e.g., Shlosman et al. 1989), but observations have not 
confirmed this. Thus, measuring the velocity at
which gas is falling from the outer parts of a galaxy towards the
center is imperative for constraining the links between large-scale
properties and activity in the nucleus of galaxies.
The dynamical analysis methods developed in 
Fathi et al. (2005, 2006) have shown to be useful for quantifying 
the kinematical effects of gravitational perturbations due to bars and 
spiral arms. Applying these methods to high-spatial resolution data for 
a large number of objects, delivered by \GF, will provide necessary 
details required for linking the kiloparsec-scale kinematics with the 
parsec-scale processes.

\subsection{Dynamical Masses of the Galaxies in the Virgo Cluster}
The large size and homogeneity of surveys of local galaxies (SDSS,
2dF, 2MASS) have yielded extraordinary advances in our understanding
of galaxy structure and evolution. One of the most important
realizations of galaxy formation studies is the recent detection
from SDSS and 2dF of a robust bimodality in the distribution of
galaxy properties, with a characteristic transition scale at stellar
mass $\sim 3 \times 10^{10}$ \Msol, corresponding to a halo (virial)
circular velocity V $\sim 120$ \kms\  (Kauffmann et al. 2003, 
Dekel \& Birnboim 2006). However, large surveys
like the SDSS lack measurements of the {\it dynamical mass}, since it 
must include coverage to luminosities below the fundamental galaxy 
transition scale at V $\sim 120$ \kms. 
Central to such a study, the number of galaxies per
unit volume as a function of halo circular velocity -- the circular
velocity function, N(V) -- is a robust prediction of cosmological
models that has never been {\it directly} tested (Klypin et al.
1999, Newman \& Davis 2002). The Virgo cluster is an ideal laboratory 
for analyzing, in depth, these processes using \GF\, observations.

\subsection{Kinematics of High-Redshift Galaxies}
In studying high-redshift galaxies, it is necessary to
disentangle distance effects from evolution ones (e.g., 
F{\"o}rster Schreiber et al. 2006). 
Even at low redshifts, controversy may exist on the nature and on
the history of merging, interacting (including compact groups (CG)
of galaxies) or even blue compact galaxies which are suspected to
constitute a large fraction of the primordial building blocks
leading to the present-day galaxies (e.g., Kunth \& \"Ostlin 2000, 
Amram et al. 2004, 2007). In compact groups, high spatial and spectral 
resolution data from \GF\, allow to observe that the broader \ha\ 
profiles are located in the overlapping regions between the galaxies 
within a group. At higher redshift such a  broadening would be 
interpreted as an indicator of rotating disk and this system 
could have been catalogued like a rotator instead of a merger 
(Flores et al. 2006). On the other hand, in cases when the spatial 
resolution increases, it becomes possible to address
the problem of the shape of the inner density profile in spirals
(CORE vs CUSP controversy), which remains one of the five main
further challenges to the $\Lambda$CDM theory proposed by Primack (2006).

The lack of resolution induce series of biases:
(i) the size of the galaxy is artificially enlarged but, 
in the same time, the size of the galaxy is diminished due 
to flux limitations; (ii) the inclination is lowered when the 
spatial resolution decreases;  (iii) the velocity gradient is 
lowered along the major axis while in the same time, the 
velocity gradient is enlarged toward the minor axis.
The consequence is that there is often no indication that
the maximum velocity of the rotation curve is reached, leading to
uncertainties in the TF relation determination.  In addition, the
(fine) structures within the galaxies (bars, rings, spiral arms,
bubbles, etc.) are attenuated or erased and the determination of the
other kinematical parameters (position angle, center, inclination)
becomes highly uncertain.

\section{The Real Integral Field Unit}

\GF\, is a real Integral Field Unit (IFU): for each pixel of the
IPCS one spectrum is available. Integral Field Units make use of lenslet arrays, fibre bundles, or slitlets to scan the two-dimensional area of interest for observations. \GF\ omits the intermediate constructions, and delivers one unique spectrum for each IPCS spectrum. There is no cross-talk between the \GF\ spectra, and each pixel corresponds to one spectrum, as opposed to each lenslet or fibre which in reality cover more than one pixel on the CCD. In this sense, \GF\ is the real IFU. This kind of instrument is called "ecological" where no information mixing is possible between spatial and spectral information.

\GF\, is composed of a focal
reducer, a filter wheel, a Fabry-Perot etalon, an Image Photon
Counting System (IPCS) as the detector, and a calibration arm (Neon
source). Data acquisition and instrument controls are done with a PC
computer via ethernet.

\subsection{Optical and Mechanical Design}

The Nasmyth focii of the William Herschel Telescope (WHT) offer two
optical modes: one with an optical derotator (in the UV/optical)
provides a Field of View (FOV) of 2.5' x 2.5' with a throughput of
75\%, while the other, without derotator,  gives access to a FOV of
5'x5' with no optical throughput degradation. In the design of the
optical system, two cases - one with derotator and one without -
were studied. Taking into account the numerous scientific cases
exposed above, we opted to keep the FOV (5'x5') as large a possible and
decided not to use the optical derotator which implied that a post observation
alignment of the data had to be developed (see section
\ref{sec:prepare}).

\begin{figure}[htp]
    \begin{center}  
   \includegraphics[width=17cm]{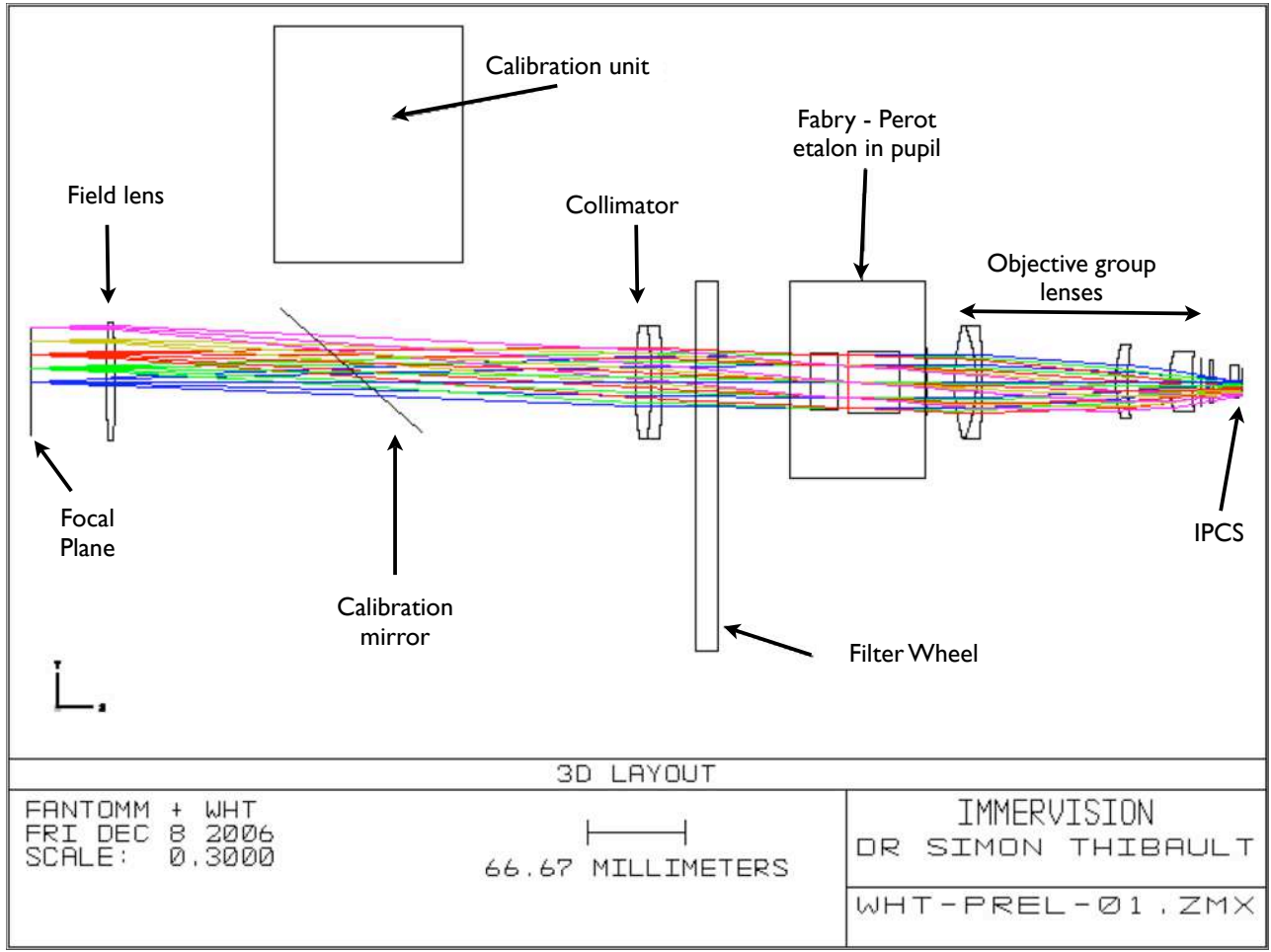}\\
      \caption{  2D optical layout of \GF. The optical design was done in collaboration with Immervision inc. Light beam from telescope comes from the left. The detector (IPCS) can been found at the right hand-side. Different colours of the optical rays represent different positions on the focal plane.  }\label{opticallayout1}
    \end{center}
\end{figure}

Optical calculations and optimisation were done using ZEMAX. 
Table \ref{pres} presents the detailed optical prescription. The optical prescription for \GF\ is beginning at surface numbered 9. Before the ninth surface, the prescription is the one the WHT for the nasmyth focus.
The focal reducer stands on the optical table of the GHRIL platform at
the 2nd Nasmyth focus of the WHT. It has a total length of 770 mm
(distance from the field lens to the detector plane). The length
between the focal plane and the field lens is 50 mm and between the
field lens and the collimator lens 345 mm. The total free space of
the collimated beam between the collimator lens and the first
meniscus of the objective is 200 mm. In the collimated beam the
pupil has a diameter of 40 mm (angles are less than  5$^o$ on the
pupil) in order to use etalon with an aperture diameter greater than
45 mm and avoid any vignetting (see Fig. \ref{opticallayout1}).

\begin{figure}[htp]
    \begin{center}  
    \includegraphics[width=17cm]{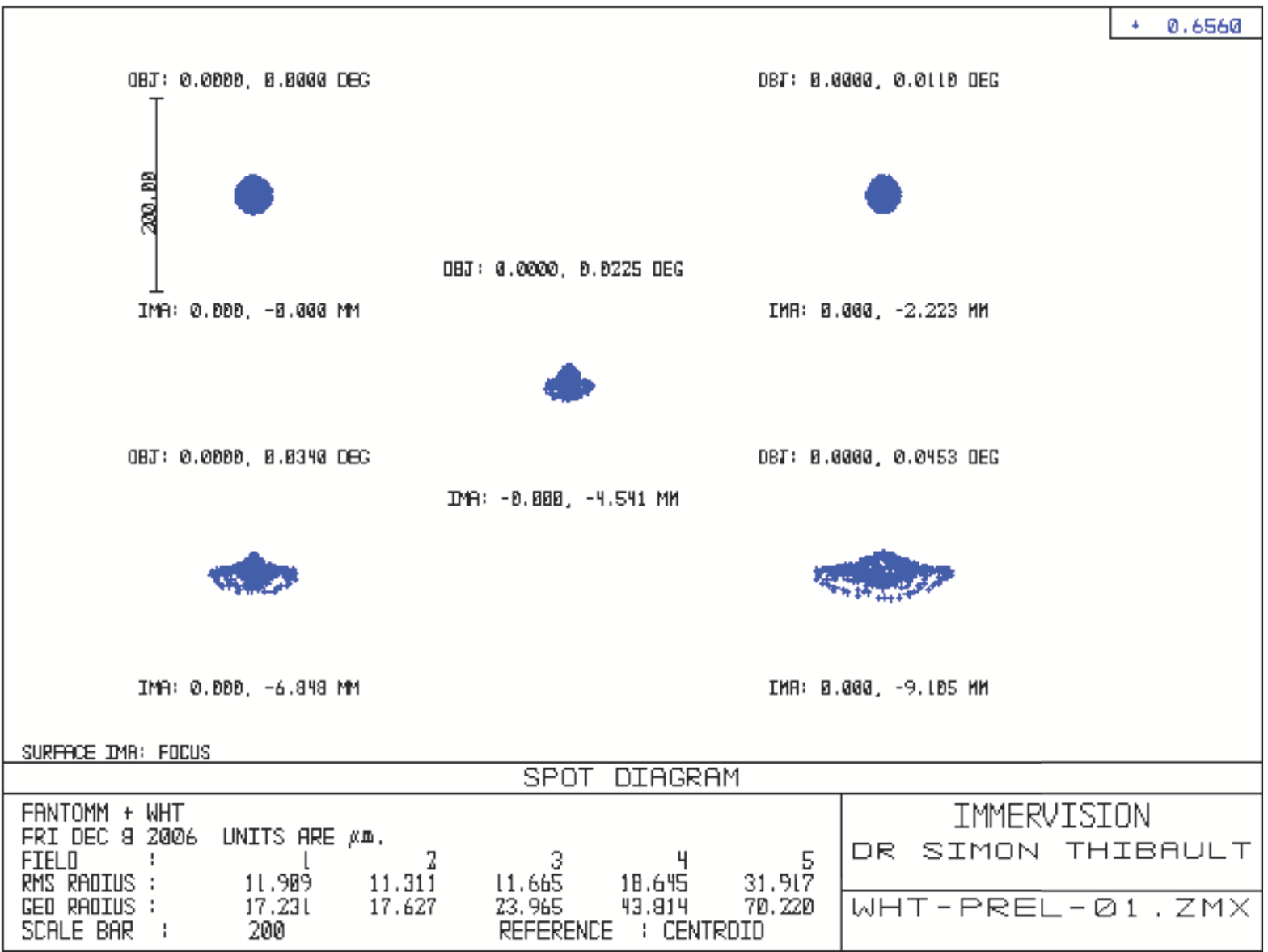}
    \caption{\label{opticallayout2}  Spot Diagram of \GF\ in the detector plane for 5 different positions of the optical rays at a wavelength of 6560\AA. }
    \end{center}
\end{figure}

Globally, as \GF\  is scanning small wavelength ranges (typically
the FWHM of a filter is 15\AA), it should be considered as an
achromatic system. The optical calculations, design and
manufacturing were relatively simple. This configuration allowed us
to use 3 commercial lenses while two custom lenses
had to be fabricated by BMV Optical Technologies \footnote{BMV
Optical Technologies 26 Concourse Gate, Ottawa, Ontario, Canada K2E
7T7}. The optical performances are shown in table \ref{tableoptique}
and the spot diagram is given in Fig. \ref{opticallayout2}.

From left to right, the optical beam will meet (a) the focal plane, where a mask has been put to avoid any diffuse light, (b) the field lens (surface \#10 in the optical prescription in table \ref{pres}), (c) the collimator lens (surface \#12), (d) a filter in the filter wheel, (e) the Fabry Perot etalon in the pupil, (f) the first objective lens (surface \#20), (g) the second objective lens (surface \#23), (g) the last objective lens (surface \#25), to be, finally, focused on the detector. Perpendicular to the optical axis, in the plane of the optical table, between the field lens and the collimator, the calibration unit (replication of the focal plane of the telescope) is founded. The calibration unit is in place when the calibration mirror is in the optical axis.

Finally, Table \ref{FP} presents the available Fabry Perot etalons
showing the ability of \GF\ to reach high ˆ resolution by
simply changing the etalon, operation that can be done in less than
2 minutes.

\begin{table}
\begin{center}
\caption{\GF\, optical characteristics}\label{tableoptique}
\begin{tabular}{p{1.5in}p{1.5in}}
\hline
Detector (Low spatial resolution mode ) & CCD, 512x512 pixels @ 25$\micron$ \\
Detector (High spatial resolution mode ) & CCD, 1kx1k pixels @ 12.5$\micron$ \\
Physical  FOV  & 12.8x12.8mm \\
FOV ( HR \& LR modes) & 202"x202" \\
FOV (HR \& LR modes on diag.) & 285" \\
Focal scale (LR mode) & 0.394"/px   \\
Focal scale (HR mode) & 0.197"/px   \\
Pupil diameter & 40 mm \\
Image Quality  &        50\% EE\footnote{Encircled energy} in 0.4" from 0 to 0.7FOV \\
            & 50\% EE $<$ 1Ó everywhere else \\
Spectral range &    450 to 850 nm \\
Geometric distortion &      $<$ 0.7\% \\
F/number in &       10.94\\
F/number out &      2.75\\
\hline
\end{tabular}
\end{center}
\end{table}

\begin{table}
\begin{center}
\caption{\GF\, optical prescription.}\label{pres}
\begin{tabular}{lllcclcc}
\hline
Surf    & Type &             Comment     &    Radius  &    Thickness           &     Glass     &  Diameter        &   Conic \\
\hline
OBJ	 & STD$^a$        &     &             Infinity    &   Infinity          &                        & 0      &        0  \\
  1 	& STD        &    &               Infinity     &       400       &                     & 4193.45      &        0 \\
  2 	& STD        &     &              Infinity      &        0        &                       & 4192.817        &      0 \\
  3 	& STD         & &                    Infinity    &      7994.4      &                  &    4192.817     &         0\\
 4 & STD            & &                Infinity          & 105.9         &                 & 4180.168   &           0 \\
5 -STO & STD  &     primary  mirror &       -20879 &     -8034.961 &              mirror  &    4180.001       &      -1\\
  6 & STD    &  secondary mirror   &    -6231.38       & 9969.75     &          mirror  &    971.8717 &      -2.53287\\
  7& STD    &                    &    Infinity    &          84    &                    &           0        &       0\\
 8 &STD    &                    &    Infinity    &         481    &        &                 113.7209      &        0\\
 9 &STD      &                    &  Infinity  &      50.39427  &                     &      72.51192      &         0\\
 10 &STD  &           011-4780      &     259.5     &       7.9         &          BK7      &  77.6889   &            0\\
 11 &STD    &                      &   Infinity     &        350     &          &             77.49541    &           0\\
 12 &STD      &        PAC096      &    258.954        &      11       &            BK7     &  56.79739         &      0\\
 13 &STD   &                         & -256.371        &     6.5             &      SF5   &   55.89834        &      0\\
 14 &STD    &                      &  -1092.828      &       100         &          &         55.25999          &    0\\
 15 &STD    &        top plate   &     Infinity        &   19.05        &      F$_{silica}$   &   37.92693      &        0\\
 16 &STD    &                     &    Infinity        &       7                &       &     35.88337          &    0\\
 17 &STD   &      bottom plate     &   Infinity        &      35      &        F$_{silica}$  &    36.86071        &      0\\
 18 &STD    &                       &  Infinity          &    18    &                    &     41.64935            &  0\\
 19 &STD   &    Distance (fix) &     Infinity    &    18.54656      &            &          45.24545    &          0\\
 20 &STD   &            PAC095     &     140.89    &        14.5              &     BK7&      49.38644 &             0\\
 21 &STD    &                    &      -101.72     &          4             &       F4     &  49.47157          &    0\\
 22 &STD   &                      &     -465.06       &       90               &     &        49.71964          &    0\\
 23 &STD    &         Custom 1     &   87.01528      &        10     &              SF6    &  49.44784       &       0\\
 24 &STD    &                       &  182.7571  &      30.15954           &       &          47.36752    &          0\\
 25 &STD   &          Custom 2    &    57.38259     &         10          &        SF11   &     38.995       &       0\\
 26 &STD    &                   &      115.5478       &        6                &      &      35.35846    &          0\\
27 &STD     &                      &  Infinity      &      6.35                   &        &  32.6863        &      0\\
28 &STD     &                   &     Infinity      &         2           &   Sapphire       &29.02896        &      0\\
 29 &STD    &                   &     Infinity      &     11.85           &      &            28.39365      &         0\\
 30 &STD    &                    &     Infinity      &       5.5  & 1.47  &  21.56853       &       0\\
 31 &STD  &                      &     Infinity       &        0                    &       & 19.45602     &           0\\
 32 &STD  &                       &    Infinity       &        2                      &     & 19.45602          &     0\\
 33 &TS$^{b}$   &        DETECTOR       &        -        &       0            &        &        18.3041        &       -\\
34- IMA& STD   &             FOCUS       & Infinity      &                       &     &           18.3041    &           0\\
\hline
\end{tabular}
\tablecomments{ a - : STANDARD, b - TILTSURF.}
\end{center}
\end{table}

The spot diagram (Fig. \ref{opticallayout2}) shows the optical performances of the instrument. It clearly stresses the size of the optical spot on the detector in various positions (conjugated of various positions on the sky). It is given for only a wavelength (6560\AA) as observations are typically done in \ha mode. The spot size is sampled by 1 pixel in the center and in alsmost the detector whereas 2 pixel are needed to have 50\% of the encircled energy on the edges of the detector. The instrument is able to well sampled the seeing (typically 0.6"/px  to 0.8"/px) with 2 pixels.

\begin{table}
\begin{center}
\caption{\GF\, Fabry-Perot etalons characteristics at
\has.}\label{FP}
\begin{tabular}{lccccc}
\hline
Etalon & p$^{a}$ & FSR$^{b}$ & $\Im^{c}$ & $\Im^{d}_e$ &R$^{e}$ \\
name & & \kms & & \\
\hline
M\'egantic 1    & 765   & 391.9 & 22 & 17 &16065\\
OM2         & 798   & 375.7 & 25 & 12 &19950 \\
OM3         & 2604 & 115.1 & 15 & 9 &39060 \\
OM4         & 609   & 492.3 & 30 & 24 &18270 \\
OM5         & 1938 & 154.7 & 56 & 32 &108528 \\
CFHT        & 1162 & 258.0 & 17 & 12 &19754 \\
\hline
\end{tabular}
\tablecomments{a - interference order. b - Free Spectral Range. c -
Finesse for a pupil point. d - Effective finesse. e - Spectral resolution.}
\end{center}
\end{table}

\begin{figure}
\centering
\includegraphics[width=1\textwidth]{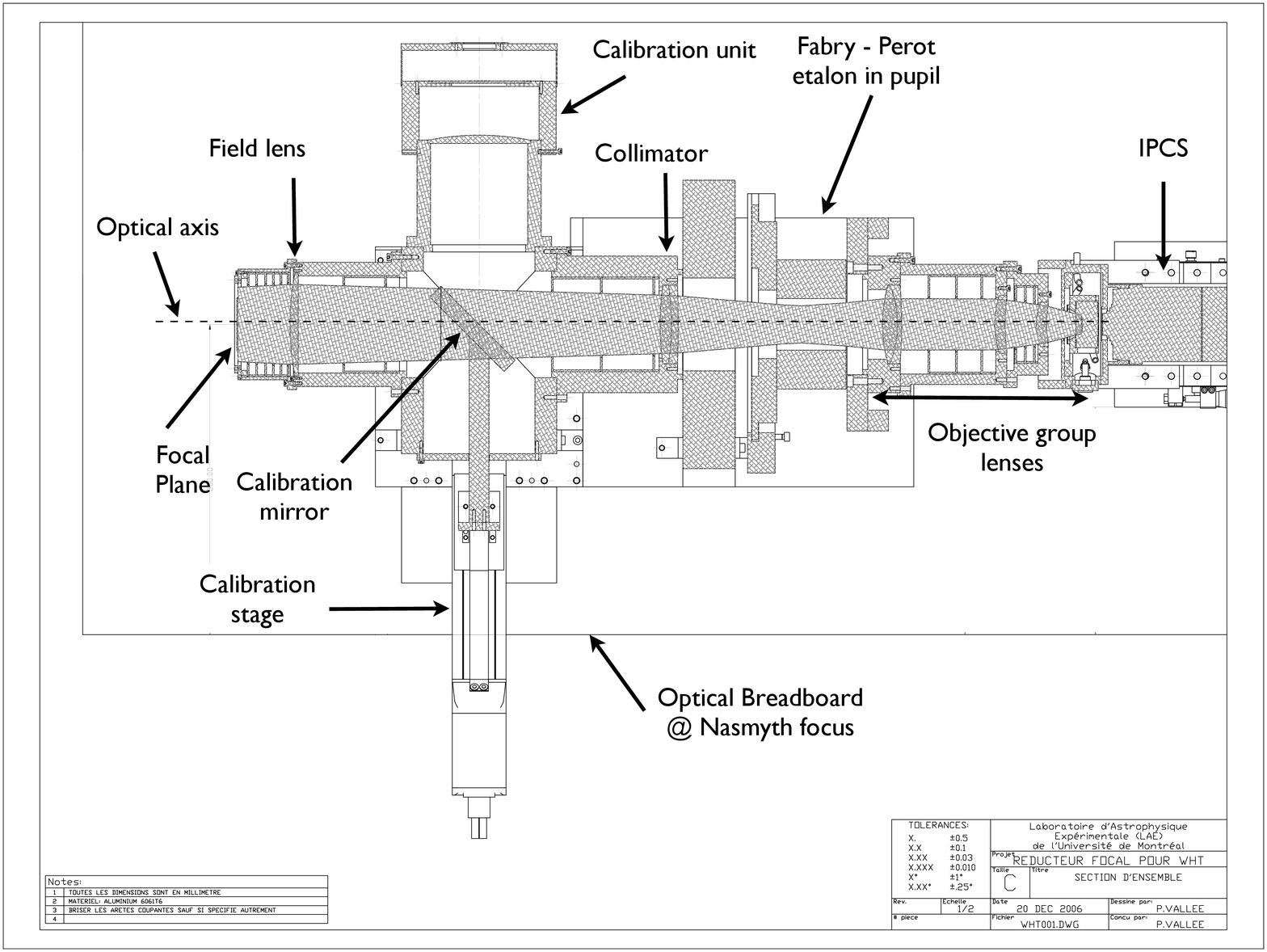}
\caption{\label{meca}Top view and cut along the optical axis parallel to the optical breadboard at the Nasmyth focus. The optical and mechanical design of \GF\ has been especially customized for the GHRIL Nasmyth focus of the WHT.
The filter wheel can fit four 75 mm-size filters and like all the
elements of the system can be controlled remotely via ethernet
cables.}
\end{figure}

The mechanical design was performed at the Universit\'e de
Montr\'eal. All parts are made of aluminium.  The lens' cylinders
were painted inside to avoid any parasite reflection and included
different masks. Lenses have been tested with a Zygo interferometer
to check their quality. The optical and mechanical integration were
done at Montreal, whereas the IPCS / Focal reducer integration was
done at Marseille. Figure (Fig. \ref{meca}) 
presents a top view and a cut along the optical axis parallel to the optical breadboard of the instrument on the optical table of the GHRIL.

Finally, the calibration arm, with its mirror controlled arm, has a
neon lamp for calibration purpose. The Ne [6598] emission is used to
calibrate the phase shifts of the Fabry Perot etalon. This lamp will
be replaced, at a later stage, by a Photon etc Tunable Source (see
www.photonetc.com) which is able to provide any calibration
wavelength.

\GF\ is fully automated and controlled via ethernet modules. Table \ref{MV} presents a summary of all automated items in their operation modes. The filter wheel, the shutter, and the calibration arms have their own ethernet modules designed by Jean-Luc Gach and Philippe Ballard from LAM. The IPCS has its own dedicated link and network to avoid a mixing of the information between ethernet modules and data. The etalon is controlled via  a RS-232 link with its CS100 controller.

\begin{table}
\begin{center}
\caption{Summary of the \GF\ automated parts}\label{MV}
\begin{tabular}{lll}
\hline
Part name & Controlled by :  & Operation  \\
\hline
Calibration arm & ethernet module & calibration \\
Calibration lamp & ethernet module & calibration \\
Calibration mirror & ethernet module & calibration \\
Calibration stage & ethernet module & calibration \\
Fabry Perot Etalon & CS 100 via RS 232 & observation \& calibration \\
Filter wheel & ethernet module & observation \& calibration \\
\hline
\hline
\end{tabular}
\end{center}
\end{table}

\subsection{Detector: state of the art IPCS}

\subsubsection{IPCS}

The IPCS is composed of a photocathode placed in front of a
commercial CCD which amplifies the photons (from a factor 10$^5$ to
10$^7$, (see Gach et al. (2002) and Hernandez et al. (2003), and
references therein for a complete description and history of such
devices). It provides a system where the detection of a photon is
around 10$^6$ times above the readout noise of the commercial CCD.
Therefore, the IPCS has essentially no readout noise.

\begin{figure}[htp]
\begin{center}
\includegraphics[width=0.30\textwidth]{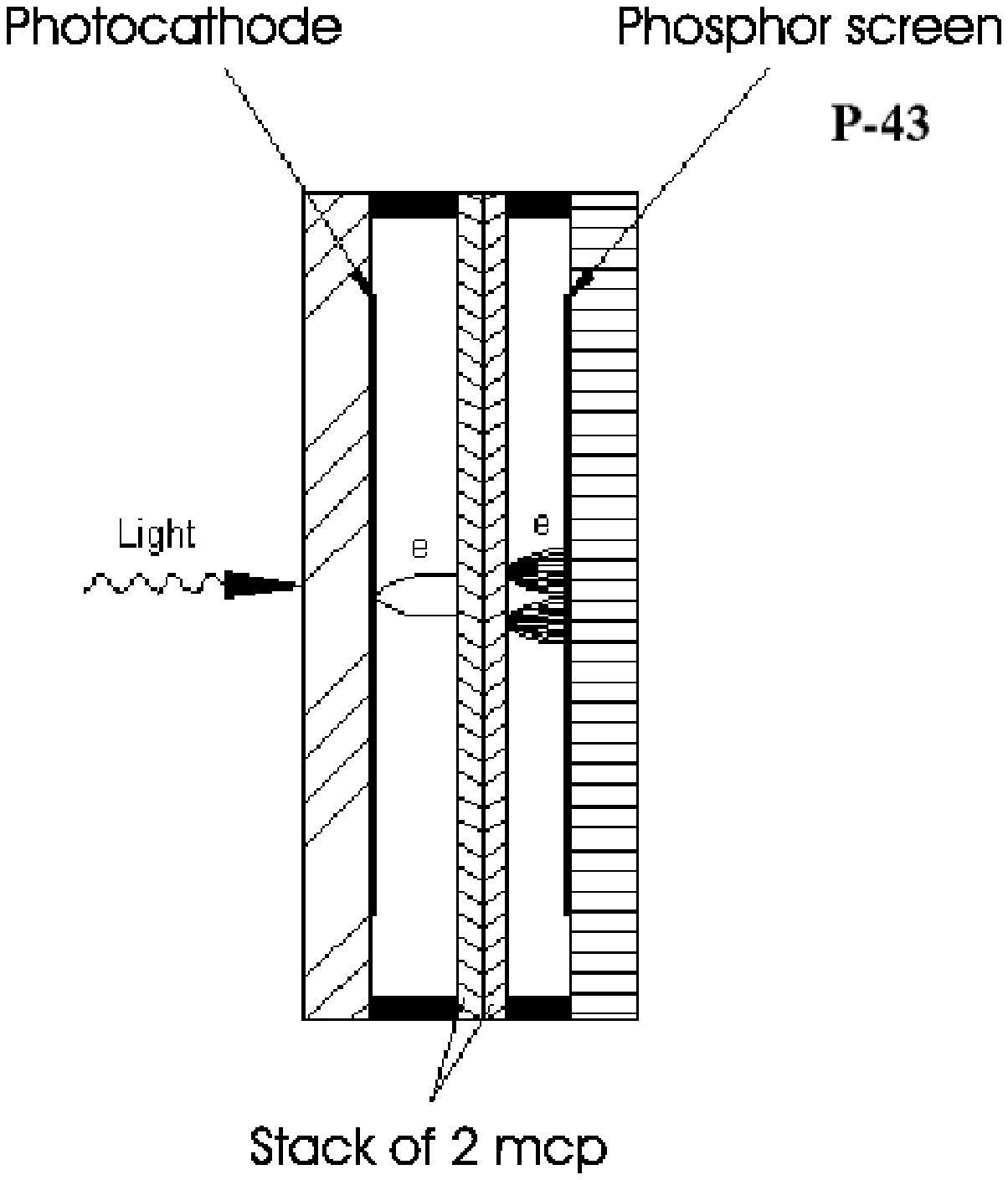}
\includegraphics[width=0.60\textwidth]{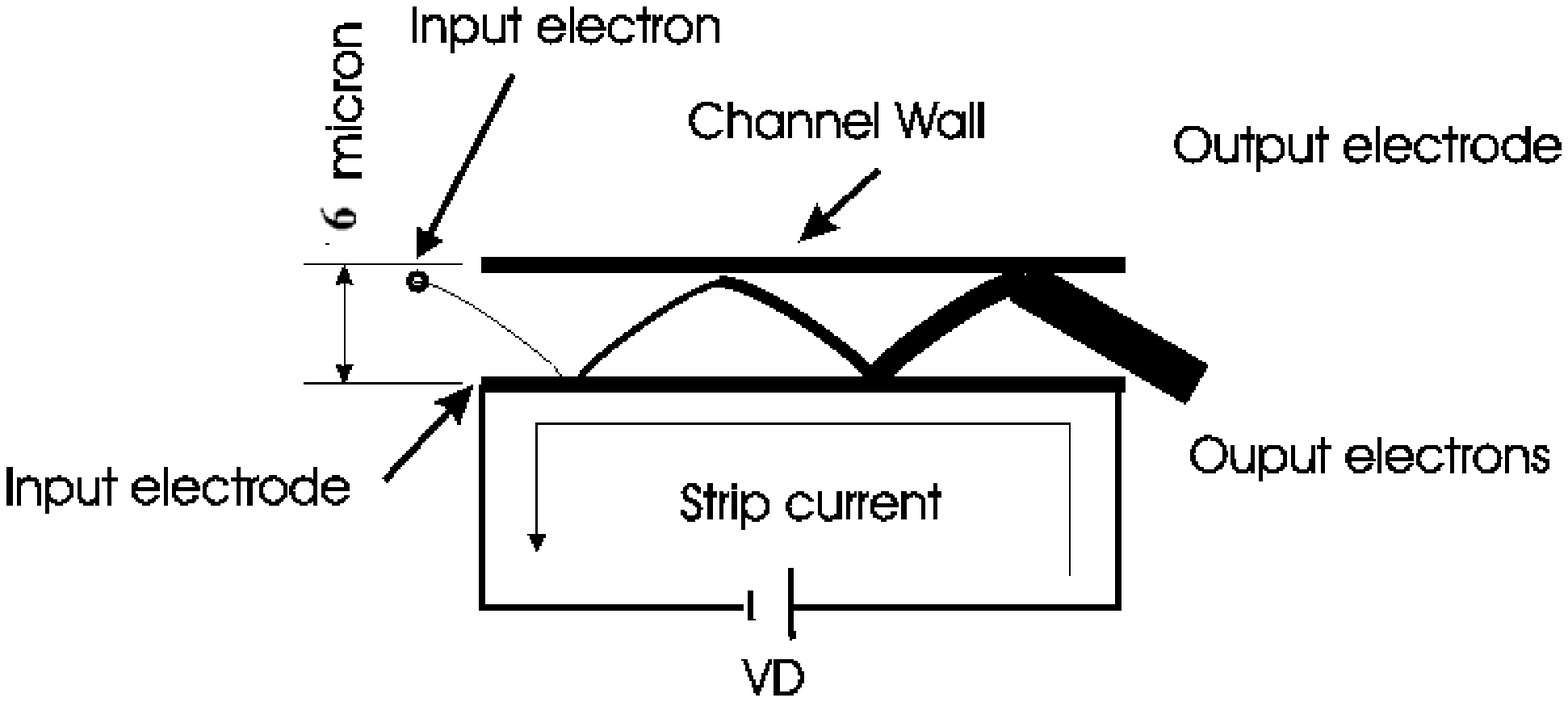}
\caption{\label{figAmpli}\textbf{Left}: Principle of electron
amplification using micro channel plates (MCP) in an image
intensifier. \textbf{Right:} Details of high voltage effect on a
micro channel.}
\end{center}
\end{figure}

The photocathode in the WHT camera is a GaAS type HAMAMATSU V7090-61\footnote{Useful documentation can be found directly on this web
site : http://www.astro.umontreal.ca/fantomm/Description/v7090u.pdf
.} with a quantum efficiency of $<\sim$26\% (this value includes the
loss due to the MCP stack). The spectral range is from 400nm to
850nm and it is very flat over these wavelengths. It uses two
microchannel plates to provide an electron amplification (after the
photoelectic effect that converts photons to electrons) of 10$^7$
(see figure \ref{figAmpli}). Each channel is 6~$\mu m$ wide.

Finally, the electrons are projected onto a P-43 phosphor screen via
a bundle of optical fibers and then is imaged on the focal plane of
a DALSA fast readout commercial CCD (Panthera 1M60) using two
objectives (from LINOS technology). Data are then transferred to an
acquisition computer using the "camera link" cable via an optical
hub and to a dedicated ethernet link at 1Gbits/s (Balard et al. 2006).

\subsubsection{Cooling System : the PELTIER Effect Simplified}

A simple commercial PELTIER effect system is used to cool down the
photocathode to its optimal temperature which is  \( -30^{o} \)C.
This PELTIER effect system is also coupled with a liquid cooling
system to evacuate heat from the IPCS. The Dalsa CCD do not need to
be cooled as photons (electrons) are amplified and well above the
readout noise. The mean thermal noise of the cathode is 5
events/frame at \( -10^{o} \)C, which is equivalent to 0.7
photon/hour/pixel and drops to 3 events/frame at \( -30^{o} \)C.
This dark current should drop to 0.4 events/frame if it was due only
to the thermal emissivity of the photocathode. This background
noise, which shows no dependence on temperature, can be explained by
K\(^{40}\) beta disintegration, which is present in the MCP glass
(Siegmund et al., 1988). Recent investigations carried out in collaboration with Biospace lab company\footnote{www.biospace.fr/} showed that using intensifier input windows free of potassium decreases the minimal background noise achievable by a factor of 10.

\subsection{IPCS Advantages}

The read-out noise affects CCDs at low light levels when very faint
flux objects are observed. Classical CCDs reach nowadays high
quantum efficiency (up to 95\%), and very low read-out noise (down
to 2-3 electrons and may be 1 in a near future,
\cite{2003PASP..115.1068G}).  With the IPCS, observations of faint
flux objects are splitted into several short ones (in particular,
this method also helps  not to be severely polluted by cosmic rays).
Thus, the maximal signal-to-noise ratio (SNR) per pixel achievable
by an ''ideal'' CCD is described by :
\begin{equation}
\label{snripcs4}
 SNR=\frac{N}{\sqrt{N+n\sigma ^{2}+T}},
\end{equation}
\noindent where \( N \) is the number of photons collected per pixel
during the exposure time, \( \sigma  \) the readout noise of the CCD
in electrons and \( T \) the thermal noise in electrons/pixel.
Although the \( T \) term is close to zero in more recent CCD or
IPCS devices and can be neglected, additional limitations constrain
the usage of CCDs.  \( n \) is the number of exposures. The \( SNR
\) decreases dramatically when \( N \) is small and \( n \) is
large. This is the case in multiplex instruments (scanning
instruments). For an IPCS, the S/N is described by the following
formula:
\begin{equation}
\label{snripcs3}
 SNR=\frac{N}{\sqrt{N+T}}\approx \sqrt{N}.
\end{equation}

As IPCS, in fact, do not have readout noise, they are also less
affected by cosmic rays since one event is seen as one photon only,
a decisive advantage with respect to CCDs, where long exposures are
required in the case of faint objects. Although the first generation
of IPCS had a poorer quantum efficiency with respect to CCDs and
were affected by image distortion, they were still competitive in
multiplex instruments or in speckle interferometry because of the
absence of read-out noise.

Multiplex instruments collect a large number of images. For example,
the scanning Fabry--Perot \FM\, uses up to 64 channels to
reconstruct the interferometric map of an object emission line, in
order to determine its velocity field, monochromatic image, and
velocity dispersion maps (see Hernandez et et al., 2003). The global
SNR of a multiplex observation (\( SNR_{m} \)) is  the quadratic sum
of the SNRs of each channel, since there is no noise correlation
between channels. Neglecting thermal noise , the SNR$_{m}$ is given
by :
\begin{equation}
\label{snripcs2}
 SNR_{m}=\sqrt{n\left(
\frac{\frac{N}{n}}{\sqrt{\frac{N}{n}+\sigma ^{2}}}\right)
^{2}}=\frac{N}{\sqrt{N+n\sigma ^{2}}},
\end{equation}
\noindent where \( N \) is the number of photons expected during the
whole exposure, \( n \) is the number of channels and \( \sigma  \)
is the readout noise of the CCD.

During multiplex observations, the emission usually appears only in a
few channels, which could be different for each pixel according to
the Doppler shift. Experiences with \textit{CIGALE} and \FM\, showed
that, emission lines appear most of the time in only 25\% of the
channels. This lowers the above \char`\"{}ideal\char`\"{} \(
SNR_{m} \). The ''worst'' case is when the line is detected only
in 2 channels. This does not affect the comparison between CCD and
IPCS usage, but gives a more precise idea of the kind of objects
that could be observed with such instruments. In this ''worst''
case the \( SNR_{m} \) is given by:

\begin{equation}
\label{ccdsnr}
SNR_{m}=\sqrt{2\left( \frac{\frac{N}{n}}{\sqrt{\frac{N}{n}+\sigma ^{2}}}\right) ^{2}}=\frac{\frac{2N}{n}}{\sqrt{\frac{2N}{n}+2\sigma ^{2}}}.
\end{equation}

Because of the very small readout time (10 ms with the WHT camera
and 25ms with the \FM\, one) in the case of an IPCS, it is possible
to observe each channel several times during the observation, then
averaging the sky variations (since the first and last images may not be comparable in terms of seeing and transparency when observing with a groundbased
telescope). Typically, each channel is observed
5 to 15 seconds, and when the last channel has been integrated, the
first is observed again. Each set of \( n \) channels is called a
cycle, the duration of which is typically 5 to 15 minutes (depending
on the exposure time per channel and the number of channels).  
The cycle exposure time must be under the OH timescale variations (typically 15 minutes, see Ramsay et al. 1992). Individual cycles are then summed to obtain the S/N ratio required without any loss of events. Many short cycles are then preferred to a single cycle with sky variation. The overlay time of the data acquisition system is under 10ms, so an exposure time of, at least, 1~s per channel, e.g. 48~s per cycle, is the shortest optimal exposure time. The experience of the LAE and LAM teams lead channel exposure time to 5~s, so a cycle time of around 5 minutes. The total observation consists of several cycles. SNR$_m$ can be
rewritten as :

\begin{figure}[htp]
\begin{center}
\includegraphics[width=1\textwidth]{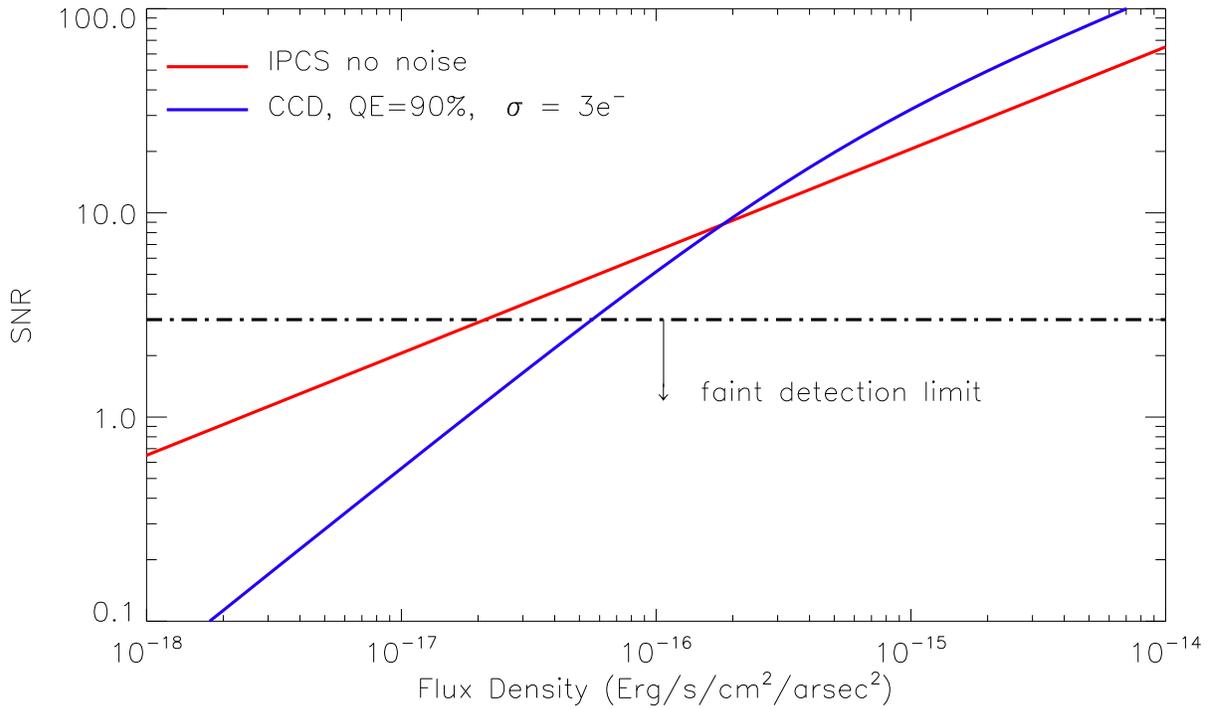}
\caption{\label{figSNRm}Signal to noise ratio comparison: IPCS and
CCD at the WHT. The red line shows the SNR for an IPCS, for a 3 hour
exposure of 40 cycles with 48 channels, on a 4.2m telescope with a pixel size of 0.5" assuming a trough-put of the telescope of 80\%. The blue line shows the SNR of
a scientific grade CCD with a QE of 90\% and a readout noise
($\sigma$) of 3$e^-$ in the same exposure conditions. The faint
detection limit is placed at a SNR of 3. }
\end{center}
\end{figure}

\begin{equation}
\label{snripcs7}
SNR_{m}=\frac{\frac{2N}{n}}{\sqrt{\frac{2N}{n}+2m\sigma ^{2}}},
\end{equation}
\noindent where \( N \) is the number of photons expected during the
whole exposure, \( n \) the number of channels, \( m \) the number
of cycles and \( \sigma  \) the readout noise of the CCD. Meanwhile,
the IPCS \( SNR_{m} \)obtained is independent of the number of
cycles and the atmosphere-free SNR is:
\begin{equation}
\label{snripcs}
SNR_{m}=\sqrt{\frac{2N}{n}}.
\end{equation}
These two equations are used to generate figure  \ref{figSNRm}. The
faint detection limit is for a SNR of 3. It can be seen on this
figure, that an IPCS can reach higher SNR than a classical CCDs for
very faint fluxes.

The measured QE of the whole system (including MCP losses) was
measured to be 26\% at H\( \alpha  \). The overall system presents
the same characteristics as the IPCS of \FM\, (Gach et al. (2002)
and Hernandez et al. (2003)). The system is limited by the process
itself. The Dalsa camera can be read with two speeds depending on
the binning mode (1024x1024px or 512x512px) chosen. When two events
occur within the same frame (16.6 ms or 10 ms ) at the same
location, only one event is counted instead of two. The mean number
M of missed photons can be evaluated assuming a Poissonian process
for the photon emission by the following equation :
\begin{equation}
\label{lin34} M=N\times(1-\frac{1-e^{-\lambda }}{\lambda }),
\end{equation}
\noindent where $ \lambda = N /${\it{frame per second}}, is the mean
number of photons expected during the resolved period of the
detector and $N$ the total number of photons.

\begin{figure}[htp]
\begin{center}
\includegraphics[width=0.8\textwidth]{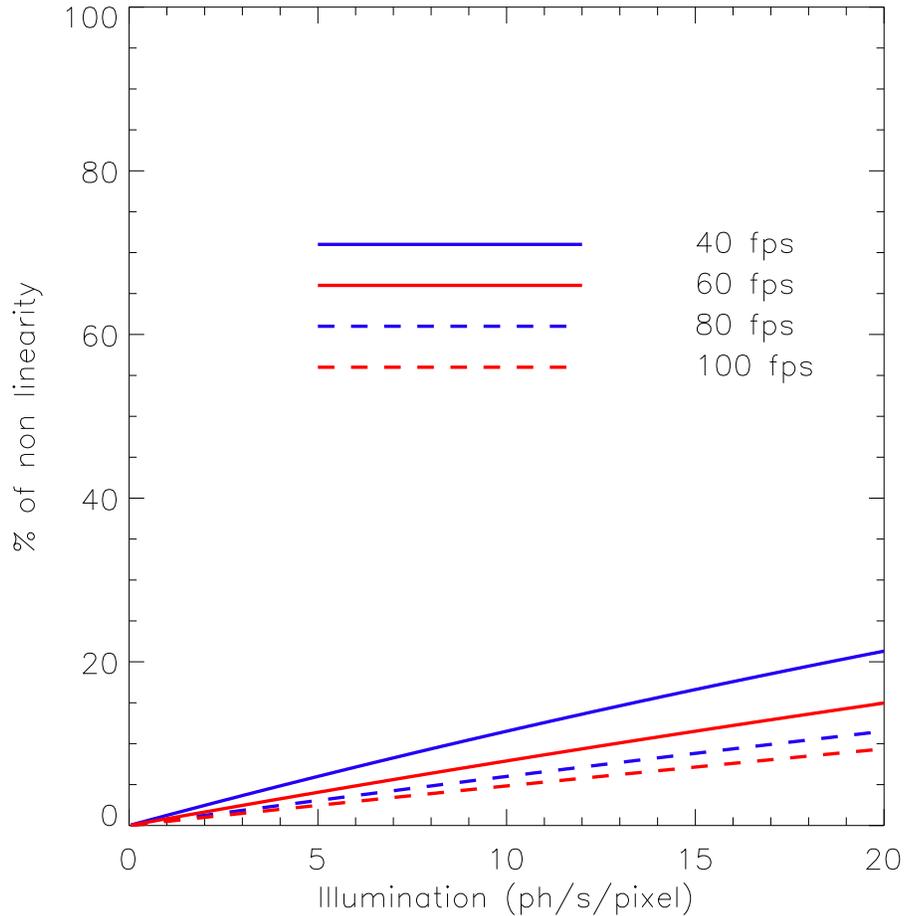}
\end{center}
\caption{\label{linear}Computed non-linearity of an IPCS detector at
different frame rates.}
\end{figure}

Figure \ref{linear} shows the percentage of non-linearity of an IPCS detector
at different possible frame rates. This shows that the system loses
its linearity when the flux increases. Therefore this detector is
used exclusively for faint fluxes. Figure \ref{linear} shows also
that this effect is less important for higher frame rates. The blue
lines (dot or continuous) represent the \FM\, GaAs system, and the
red one is the improved \GF\, camera.

\section{Data Reduction}
\subsection{Data Preparation}
\label{sec:prepare}

As \GF\ is mounted on the optical table of the Nasmyth focus, our
observations are affected by the field rotation. The optical
derotator provided by the Isaac Newton Group (ING) of Telescopes has
the disadvantage of covering a field of only $2.5\arcmin \times
2.5\arcmin$, which is too small for the scientific programs
envisaged for \GF. Our observation strategy allows us to store the
image of the individual channel separately, which is ideal for
derotating the observed cube {\em a posteriori}. Following the
information provided on the ING web pages, the field rotation is
$\approx 0.29$\arcsec, i.e., negligible, during the exposure of each
individual cycle ($\sim 4$ minutes).

Most of the data reduction steps are carried out using an IDL based
data reduction package developed by Daigle et al. (2006). However,
since we do not use the optical derotator, the data are affected by
the rotation. In order to correct for this effect, we combine
Daigle's reduction package with a number of available packages,
KARMA\footnote{http://www.atnf.csiro.au/computing/software/karma},
SWARP\footnote{http://terapix.iap.fr} and
IMWCS\footnote{http://tdc-www.harvard.edu/software/wcstools/imwcs}.
The complete steps of the \GF\, data reduction are:

\begin{enumerate}
\item An initial and rough estimate of the data quality by reducing the observed data cube. This step allows also a good phase calibration which we use to reduce the individual cycles.

\item Individual reduction of each observed cycle.

\item Creating a cube image for each reduced cycle. Investigating each of these images, we have found that for almost all the observed galaxies, the effect of the sky rotation is marginal within the 4 minutes exposure to scan one cycle. This is also confirmed by the ING web pages.

\item The collapsed cube image of each individual cycle is then used to find the three brightest point sources. The IMWCS and KARMA tools are then used to include the appropriate keywords in the header of each image.

\item We found that a simple rotational transformation does not provide the adequate accuracy, hence we use the SWARP package to match all cycles regardless of the complexity of the transformation required.

\item We apply the same transformation to each individual channel within each cycle, and re-build a corrected version of the data cube.

\item Finally, we reduce the corrected data cube following all the standard reduction procedures presented in Daigle et al. (2006).
\end{enumerate}

\begin{figure}[htp]
\begin{center}
\includegraphics[width=0.9\textwidth]{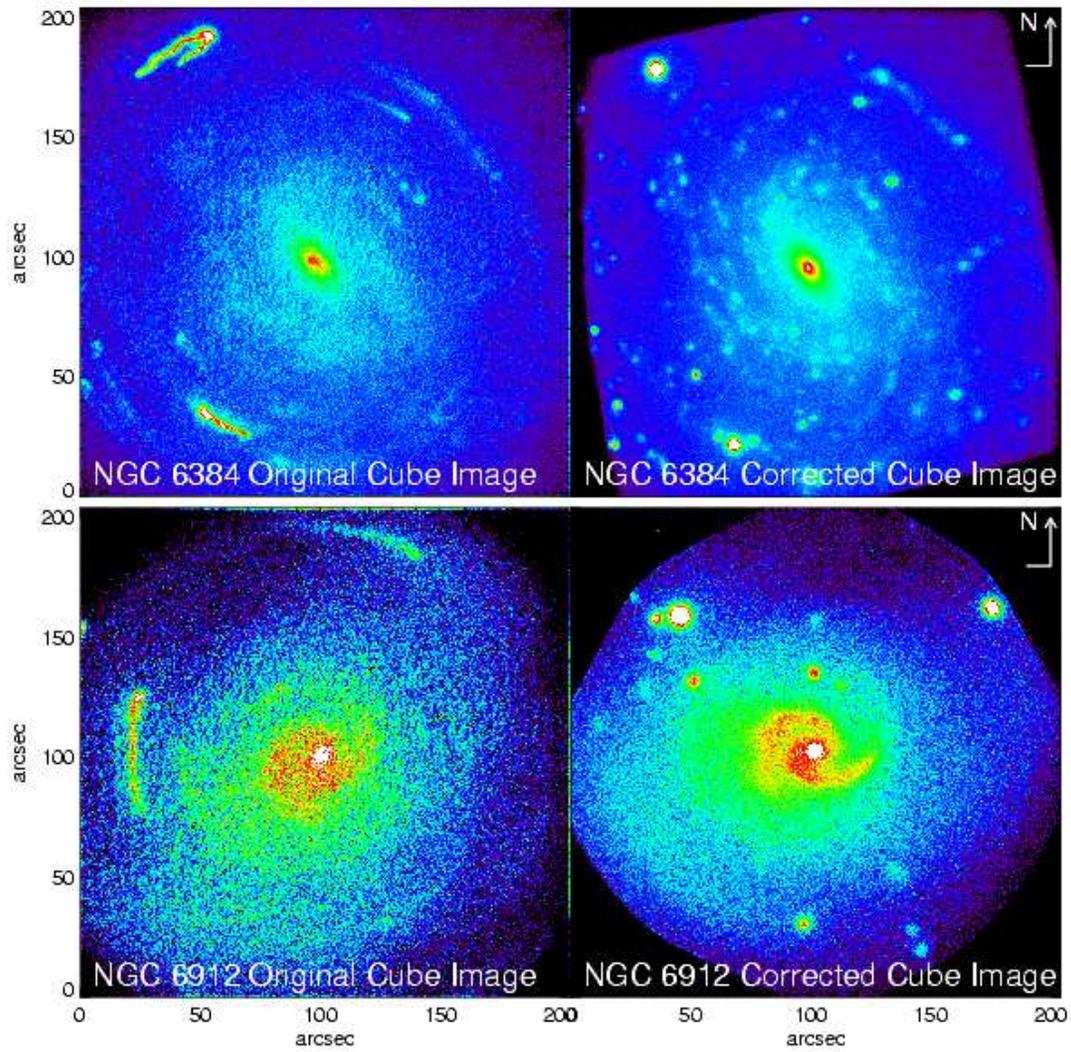}
\caption{Examples of derotation}
\end{center}
\end{figure}

Note that this process is made possible only because our detector has a high time resolution and has no readout noise. This reduction process does not add any excess noise to the observational data and keeps the very high overall sensitivity of the instrument.

\subsection{Data Reduction and Derivation of the \ha Kinematics}
\label{sec:reduction}

\GF\ data reduction involves the standard Fabry-Perot data reduction
steps: cube construction, phase calibration, sky subtraction, and
kinematics derivation, which is done using the software package of
Daigle et al. (2006) and the GIPSY package
\citep{2001ASPC..238..358V}.

By construction, \GF\, covers a reasonably large field in order to
observe nearby extended objects and provide simultaneous sky
coverage to allow for reliable sky subtraction. However, some
galaxies extend beyond the full \GF\ field. In order to sample the
sky variations properly during the long observation of the object,
the telescope is pointed to a blank field away from the galaxy, once
every three cycles. The sky cubes are thereafter treated in the same
way as the galaxy cube. The reduced sky cube is then re-scaled to
the galaxy count level and subtracted from the galaxy cube. In order
to examine the accuracy of the sky subtraction procedure, the sky
cube is used in four different ways before subtraction from the
galaxy. All the sky spectra are then rearranged to build one
spectrum representative of the entire sky cube, but also fitted with
1st, 3rd and 4th order polynomials to each channel image of the sky
cube. Subtracting each sky representation separately, we end up with
four different reduced galaxy cubes.

The free spectra range FSR determines the Ç depth È of the data cube and is a function of the observed wavelength ($\lambda$) and the interference order ($p$) at this wavelength. $FSR = \lambda / p$. The number of channels is linked to the finesse of the Airy function of the etalon (in fact, the number of channel respects the Nyquist criteria : n = 2.2 $\times$ the effective Finesse) and has no relation with the FSR. To study normal galaxies, a FSR of 0.8nm (360 kms at \has) is used. If the velocity range of a galaxy (with its inclination factor) is larger than the FSR (which is not frequent), one will find in order -1 or +1 the rest of the velocity field (filters, used in front of the etalon, typically block all etalon order except $p+1, p, p-1$). So a jump in the velocity field is generally found and easily corrected. 
This could happen with active galactic nuclei of galaxy. In that particular case, an other etalon is used to increase the FSR.  Table 2 presents the FSR for different etalons.

Each reduced galaxy cube is treated independently to derive the \ha\
kinematics by means of deriving the moment maps with consistency
checks by applying single Gaussian profiles to the spectra (e.g.,
Fathi et al. 2007a). Analyzing the kinematical maps, it was found
that the sky subtraction does not significantly change the derived
kinematics. In order to gain more field coverage, a two-dimensional
Voronoi tessellation method (Cappellari \& Copin 2003, Daigle et al
2006) was applied on the faint regions of the observed field.
Finally, the maps are cleaned by removing all values which are
derived from spectra with amplitude-over-noise ($A/N$) less than 10
(the noise being the standard deviation of the parts of the spectra
outside the \ha\ emission line).  Although this seems a strict
criterion, it was chosen to continue the analysis of the velocity
field based on spectra which are reliable not only for deriving
reliable velocities, but also velocity dispersion.  \GF\
instrumental dispersion has been measured using the Neon lamp and
was found to be $\sigma_\mathrm{inst}=FWHM/2.35=2.5$ \kms. It was
then subtracted quadratically from the \ha\ velocity dispersion, and
the derived velocities were corrected for the Heliocentric velocity
drift.

To give an idea of the efficiency of the pipeline reduction, two
panels of results are given, representing the scientific drivers of
the instrument. The final kinematical maps are presented in the appendix section  in
Fig.~\ref{fig:kinematics} for four objects while \ha monochromatic
images are represented in Fig~\ref{fig:mono}.  Finally,monochromatic image in the [NII] line and the first order of the PNe M1-75 is presented in Fig~\ref{fig:PNe}. This is not the goal
of this paper to comment, analyse or derive kinematical parameters.
The final, complete and detailed data reduction and analysis will be
available in forthcoming papers.


Finally, the status of the data reduction of the first run (17 objects
!) over 6 nights are presented in Table \ref{results}.

\begin{table*}[htp]
\begin{center}
\caption{\label{results} Observed Objects with \GF\ during 6 nights}
\begin{tabular}{lrrlccl}
\hline
Name& R.A. & Dec. & Vsys & Night & $\langle$Seeing$\rangle$ & Status\\
\hline\hline
NGC 4376    &  12 25 18.0 &  +05 44 28&  1136         & 6     &    0.7  & Reduced\\
NGC 4594    &  12 39 59.4 &  -11 37 23&  1024         & 4     &    2.0  & Reduced\footnote{Dicaire et al 2008}\\
NGC 5236    &  13 37 00.9 &  -29 51 57&  513          & 5     &    0.7  & Reduced\footnote{Fathi et al 2008}\\
NGC 5427    &  14 03 26.0 &  -06 01 51&  2618         & 2     &    1.5  & Reduced\\
NGC 5850    &  15 07 07.7 &  +01 32 39&  2556         & 3     &    2.0  & Reduced\\
NGC 5954    &  15 34 35.1 &  +15 11 54&  1959         & 2     &    1.5  & Reduced\\
NGC 6118    &  16 21 48.6 &  -02 17 00&  1573         & 6     &    0.6  & Reduced\\
NGC 6384    &  17 32 24.3 &  +07 03 37&  1665         & 1     &    1.8  & Reduced\\
NGC 6643    &  18 19 46.4 &  +74 34 06&  1484         & 2     &    1.5  & In progress\\
NGC 6643    &  18 19 46.4 &  +74 34 06&  1484         & 2     &    Imaging& In progress\\
NGC 6912    &  20 26 52.1 &  -18 37 02&  6968         & 3     &    2.5  & Reduced\\
NGC 6912    &  20 26 52.1 &  -18 37 02&  6968         & 6     &    0.8  & Reduced\\
NGC 6962    &  20 47 19.1 &  +00 19 15&  4211         & 4     &    1.8  & Reduced\\
KPG 552     &  21 07 44.0 &  +03 52 30&  7767         & 5     &    0.6  & Reduced\\
KPG 591     &  23 47 01.6 &  +29 28 17&  4845         & 4     &    1.5  & Reduced\\
PN M1-75    &  20 04 44.1 &  +31 27 24&  $\ddag$3890  & 5     &    0.8  & Reduced\\
Arp 278     &  22 19 28.6 &  +29 23 32&  4718         & 6     &    0.8  & Reduced\\ \hline
\end{tabular}
\tablecomments{All objects that were successfully observed during the commissioning run of 6 nights.  $\ddag$: distance measured in parsecs. Note that NGC~6912 was observed twice in order to achieve improved spatial resolution, and the planetary nebula, PN~M1-75, was observed in high-resolution mode with 0.2 arcsec/pix.}
\end{center}
\end{table*}

    \section{Conclusions}

\GF\ , a new Fabry-Perot system with high spatial and spectral
resolution has been presented. \GF\ offers the largest FOV of such
instrument on a 4m class telescope. It is also the most versatile
instrument capable to study ISM to high redshift galaxies.

\GF\ provides a FOV of 202"$\times$202" with two spatial
resolutions: 0.4"/px or 0.2"/px (hard-binned mode). Different
spectral resolutions can be achieved simply by changing etalon in
the pupil space. With the actual bank of etalon available,  spectral
resolution varies from 8000 to 110000. The use of tunable filters
(in development both at LAE and LAM) would also improve the
versatily of this instrument. The high performance no readout-noise
detector, in the \GF\ version, remains  the best detector to be used
with faint fluxes science cases. It allows, at the same time,
real-time data acquisition (astronomical object is seen in live
during observation), on sky reduction and real-time OH sky lines
removing. Finally, a robust reduction pipeline, for complete data
reduction is available and can be obtained at
www.astro.umontreal.ca/fantomm/reduction.

Developing of \GF\ was formulated in the summer of 2006 by the IAC
and conception, design and realization began in the beginning of
2006 December at LAE. The first lights were obtained only 6 months
later in 2007 summer. This fast schedule was possible because of th=e
expertise developed at LAM, Observatoire de Marseille, and LAE at
University of Montreal, through the building of \FM\ and
\textsc{CIGALE}. While \GF\ was constructed as a private instrument,
it is expected to become a visitor instrument at WHT in less than
two years.

\GF\ should be considered as an high-performance instrument but also
as a lab test of much more complicated instruments such as the {\sc
3D-NTT} (www.astro.umontreal.ca/3dntt), the Brazilian Tunable Filter
: \textsc{BTF}i (www.astro.iag.usp.br/$\sim$btfi) and the {\sc
smart} tunable Filter for the E-ELT (Moretto et al. 2006). Those
instruments will use \GF\ technology but with many improved
features, such as EMCCD (Daigle et al. 2007), customs large tunable
filters and new Tunable filter made with Volume Phase gratings
(Blais-Ouellette et al. 2006).

\acknowledgments

We would like to thank the staff of the WHT,  where the data were
obtained, for their continuing support. Special thanks also to
Olivier Boissin for his help before and during the commissioning of
the instrument. We also acknowledge support from the Natural
Sciences and Engineering Research Council of Canada and the Fonds
Qu\'eb\'ecois de la recherche sur la nature et les technologies.
This research has benefitted from support by the following projects:  AYA2004-28251-CO2-01 and AYA2007-67625-CO2.-01 of the Spanish Ministry of Education and Science, and P3/86 of the Instituto de Astrof'sica de Canarias. Kambiz Fathi acknowledges support from the Wenner-Gren foundations, the Royal Swedish Academy of Sciences' Hierta-Retzius foundation, and the IAC project P3/86. Monica Rela\~no acknowledges support from the Juan de la Cierva Fellowship Program and the project AYA2004-08251-C02-00. We are grateful to Daniel Fabricant, associate editor of PASP,  for helpful comments.

\facility{Based on observations obtained at the 4.2mWHT telescope operated on La Palma by the Isaac Newton Group in the Spanish Observatorio del Roque de los Muchachos of the Instituto de Astrofisica de Canarias, on the island of La Palma}

\appendix

\clearpage

    \begin{figure}[p]
    \plotone{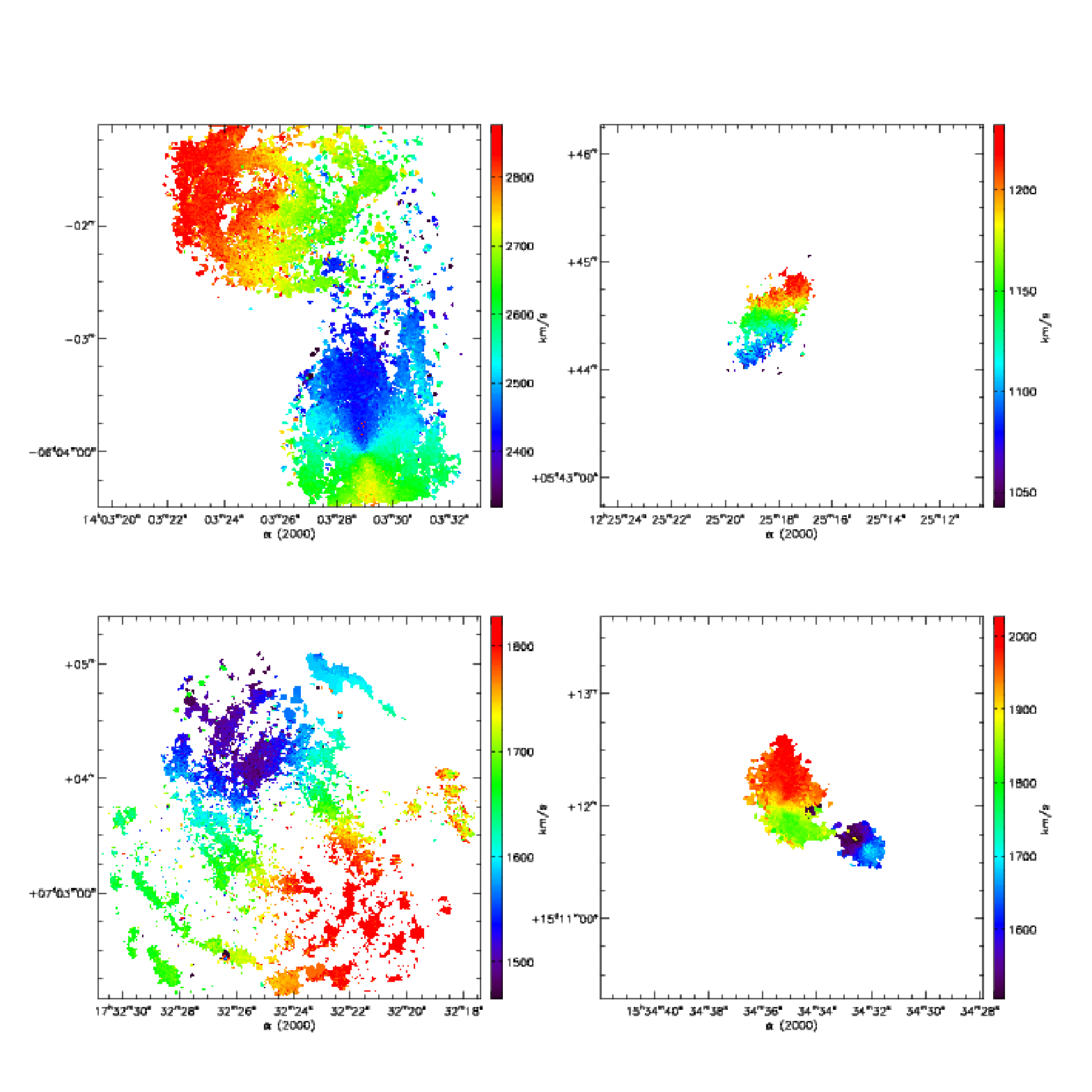}
        \caption{\label{fig:kinematics}Kinematical maps of 4 objects.{\bf Top left}: NGC 5427  with NGC 5426.{\bf Top rigth}:  NGC 4376.{\bf Bottom Left}:  NGC 6384. {\bf Botton right}: NGC 5954 with NGC 5953}
    \end{figure}

        \begin{figure}[p]
    \plotone{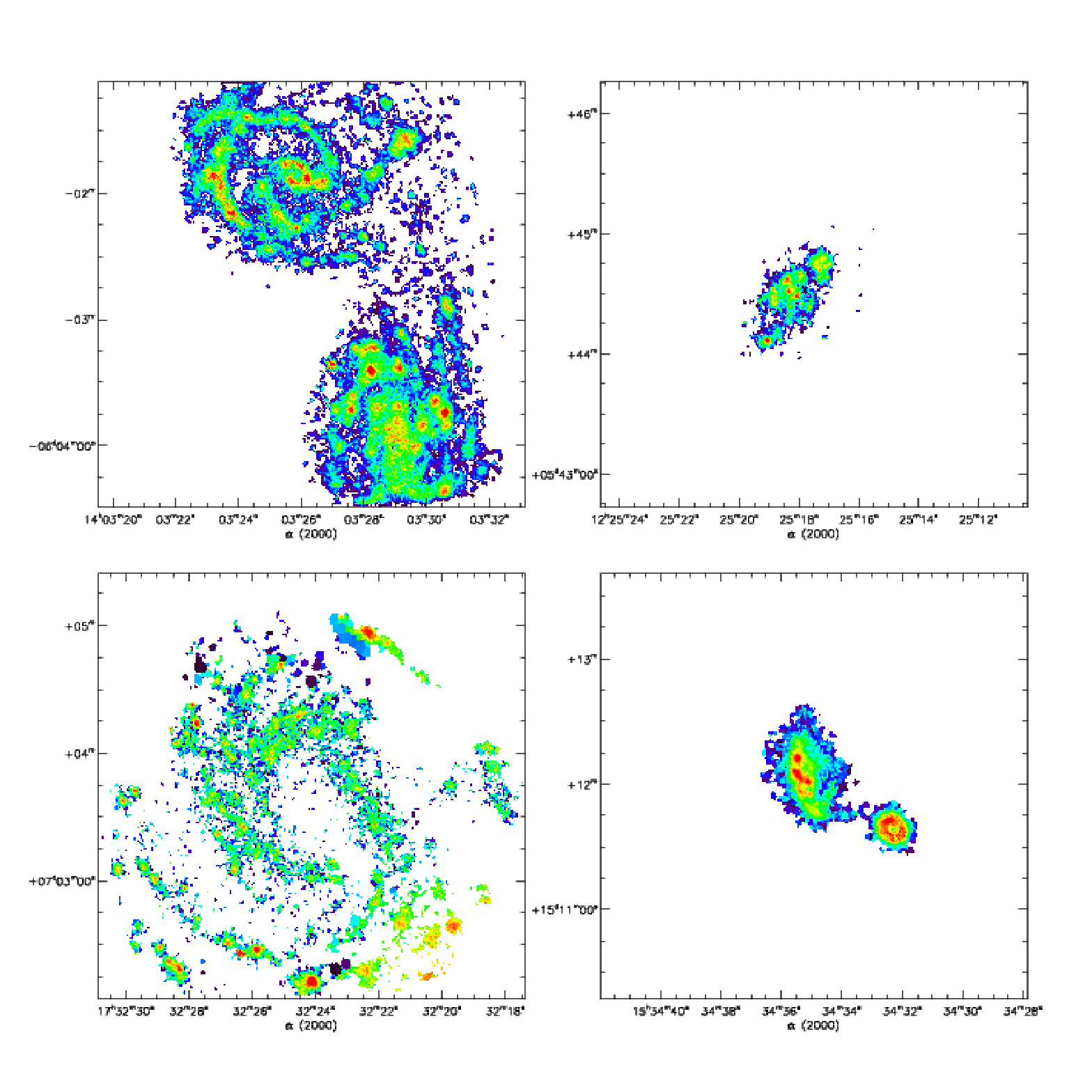}
        \caption{\label{fig:mono}\ha images of 4 objects.{\bf Top left}: NGC 5427  with NGC 5426.{\bf Top rigth}:  NGC 4376.{\bf Bottom Left}:  NGC 6384. {\bf Botton right}: NGC 5954 with NGC 5953}
        \end{figure}

   \begin{figure}[p]
  \centering
\plottwo{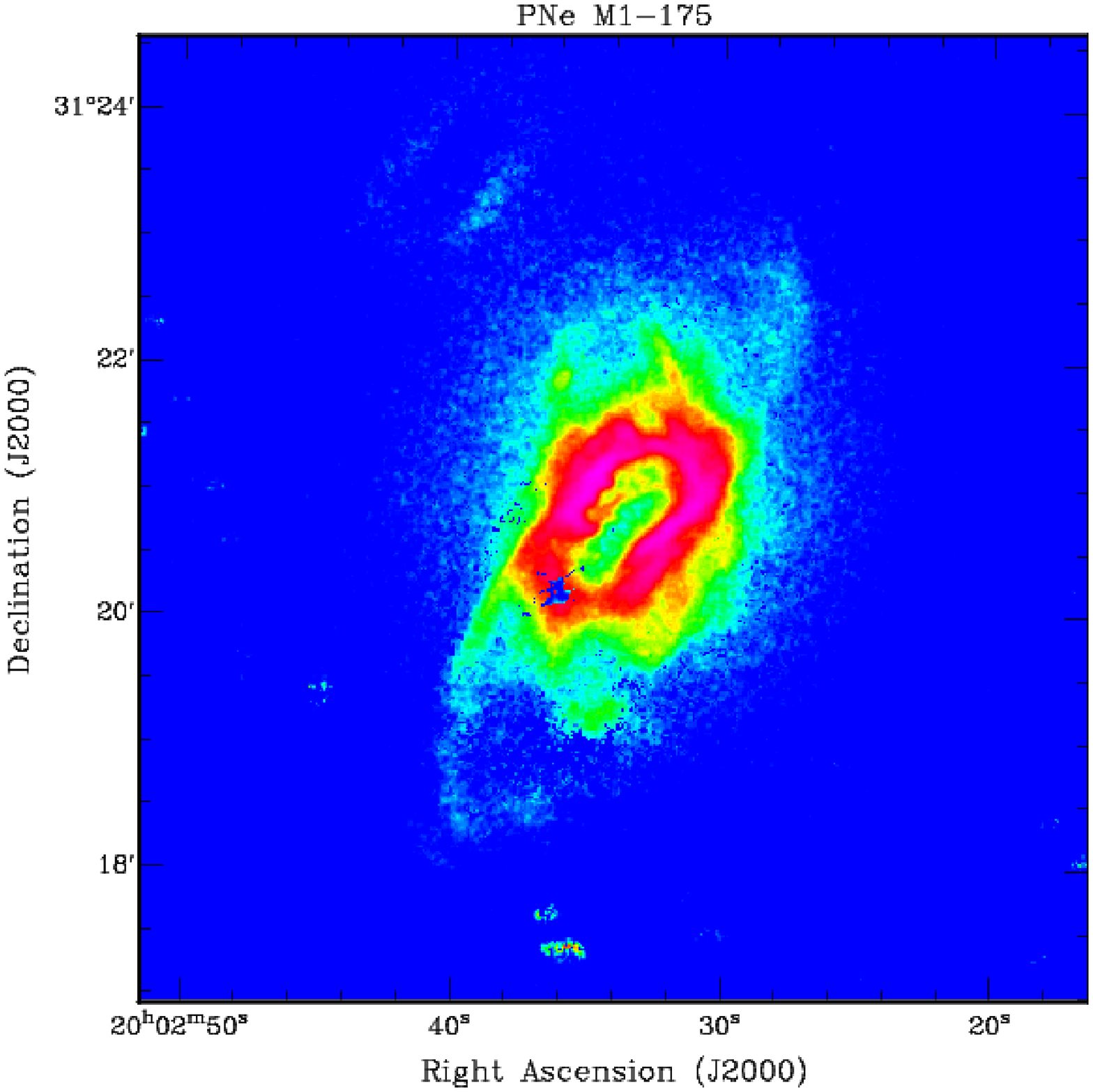}{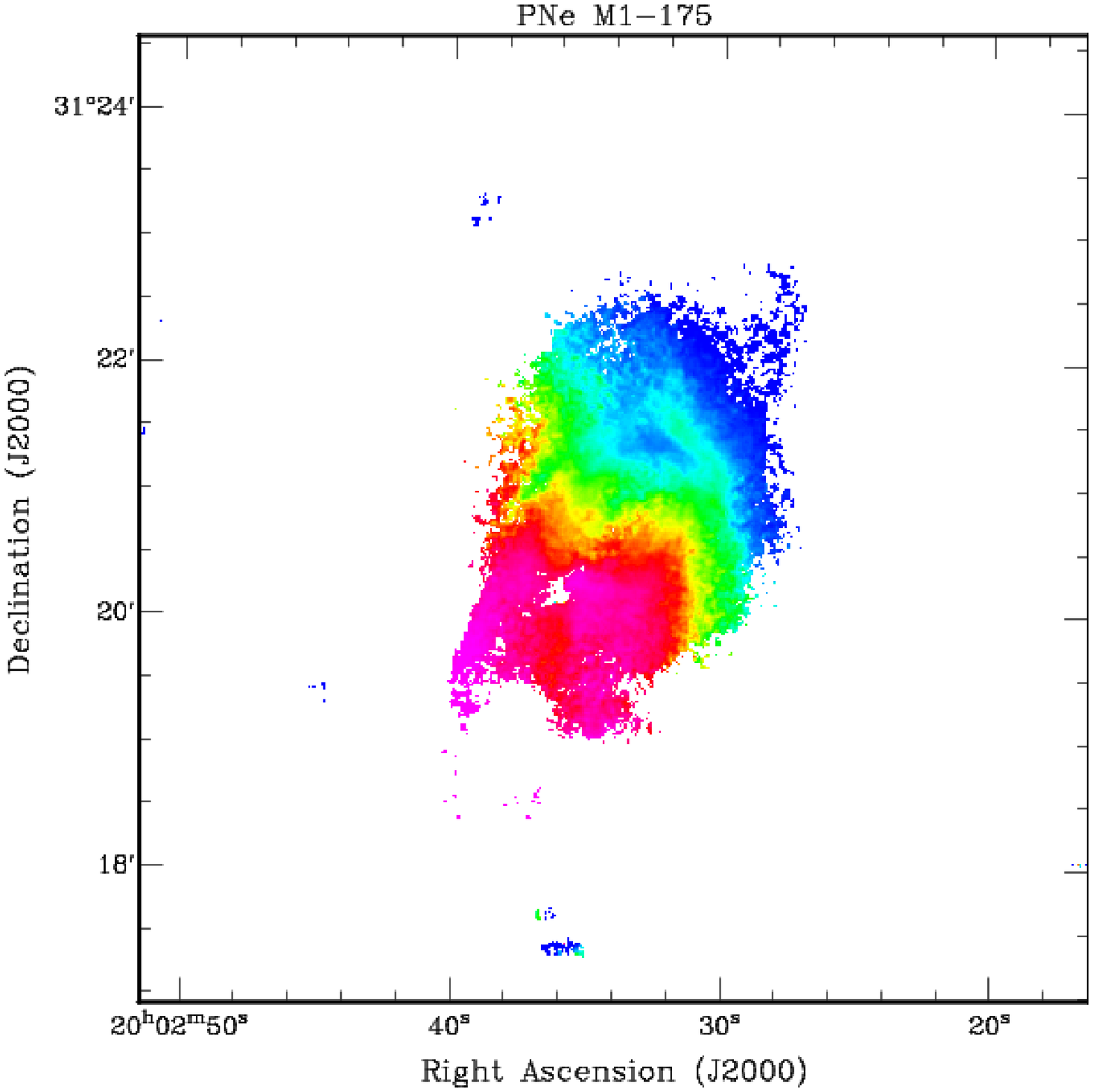}
       \caption{{\bf Left}: \label{fig:PNe}[NII]658.3~nm image of Planatary Nebula M1-75 with a pixel size of 0.2". {\bf Right}: associated expansion velocity field. Range of velocity is from 76 \kms to 231 \kms}
     \end{figure}

\end{document}